\documentclass[10pt]{article}
\usepackage{amsmath,amsthm,amsfonts}

\renewcommand{\thesection}{\Roman{section}}
\renewcommand{\thesubsection}{\Alph{subsection}}

\newtheorem*{proposition}{Proposition}

\def\e{\mbox{e}}

\def\tv{\vartheta_4}
\def\F{{\cal F}}
\def\G{{\cal G}}

\begin{document}

\title{The dilute A$_4$ model, the E$_7$ mass spectrum
and the tricritical Ising model\thanks{Expanded version of a talk
presented at the International Workshop on
Exactly Solvable Models of Statistical Mechanics and Mathematical Physics,
Asia-Pacific Center for Theoretical Physics (Seoul, Korea, June 26-29, 2000).}}

\author{K. A. Seaton\thanks{Permanent address:
School of Mathematical and Statistical Sciences,
La Trobe University, Victoria 3086, Australia.}\\
C. N. Yang Institute for Theoretical Physics,\\
State University of New York,
Stony Brook, NY 11794-3840, USA
\and
M. T. Batchelor\\
Department of Mathematics, School of Mathematical Sciences,\\
Australian National University, Canberra ACT 0200, Australia}

\date{}

\maketitle


\begin{abstract}
The exact perturbation approach is used to derive the (seven)
elementary correlation lengths and related mass gaps of
the two-dimensional dilute A$_4$ lattice model in regime $2^-$
from the Bethe Ansatz solution.
This model provides a realisation of the integrable $\phi_{(1,2)}$ perturbation
of the $c=\frac{7}{10}$ conformal field theory, which is known to describe
the off-critical thermal behaviour of the tricritical Ising model.
The E$_7$ masses predicted from purely elastic scattering theory
follow in the approach to criticality.
Universal amplitudes for the tricritical Ising model are calculated.
\end{abstract}

\newpage

\section{INTRODUCTION}\label{int}

The deep relationship between conformal field theory and criticality
has provided a wealth of detailed information on phase transitions
and critical phenomena.
Moreover, {\em perturbed} conformal field theory provides a description of the
{\em approach} to criticality in certain models \cite{Hbook}.
One of the most striking examples
is the $\phi_{(1,2)}$ perturbation of the
minimal unitary conformal field theory ${\mathcal M}_{3,4}$ which is
known to describe the scaling limit of the two-dimensional Ising
model at $T=T_c$ in a magnetic field.
In particular, Zamolodchikov's construction of nontrivial local integrals
of motion and thus an integrable quantum field theory led to the remarkable
prediction of eight fundamental mass ratios for the magnetic Ising model
\cite{Zb}.
The masses coincide with the components
of the Perron-Frobenius vector of the Cartan matrix of the Lie algebra E$_8$.

In another development, the exactly solvable dilute A$_3$ lattice model was
discovered \cite{WNS} and (in regime 2 of its four regimes) seen to be
in the same universality class as the magnetic Ising model.
Most importantly the dilute A$_L$ model \cite{WNS, R} admits an off-critical
extension in which the Boltzmann weights are parametrised in terms of elliptic
theta functions \cite{WNS}.
In the dilute A$_3$ model the elliptic nome plays the role of magnetic field.
Its hidden E$_8$ structure has been revealed by a
number of studies \cite{BNW}-\cite{ BS2}.
%
The masses, obtained from the eigenspectrum, may be
summarized by the formula \cite{MO,BS2}
\begin{equation}
m_j \sim \sum_a \sin \left( \frac{a \pi}{g} \right) ,\label{m8}
\end{equation}
where index $j$ labels the eight particles, $g=30$ is the Coxeter number
for E$_8$
and the set of allowed $a$ values is given in Table \ref{eight}.


In addition to the correspondence between the dilute A$_3$ model and E$_8$,
there
are similar correspondences between the dilute A$_4$ model and E$_7$, and
the dilute A$_6$ model and E$_6$.
In regime 2 these models are lattice realizations of the $\phi_{(1,2)}$
perturbation
of the ${\cal M}_{4,5}$ and ${\cal M}_{6,7}$ minimal unitary conformal
field theories
respectively, known to have connection to the other exceptional Lie
algebras \cite{FZ}.
Some E-type structures have been observed for these dilute A models
\cite{WPb,S2}.

Based on the results for the eigenspectrum of the dilute A$_3$ model
\cite{BS2}
and general inversion relations, we proposed \cite{BS3,SB} that, in the
thermodynamic
limit and in the appropriate regime, the row transfer matrix eigenvalue
  excitations
\begin{equation}
r_j(w)=\lim_{N\to\infty}
\frac{\Lambda_j(w)}{\Lambda_0(w)}\label{excite}
\end{equation}
of the dilute A$_3$, A$_4$ and A$_6$ models are
given by the following general expression.

\begin{proposition} The excitation spectrum of the dilute A$_3$, A$_4$ and
A$_6$ models
in regime 2 is given by
\begin{equation}r_j(w) = \prod_a w\,\frac{E(-x^{\frac{6sa}{g}}/w,
x^{12s})
E(-x^{\frac{6s(g-a)}{g}}/ w, x^{12s})}
        {E(-x^{\frac{6sa}{g}}w, x^{12s})
E(-x^{\frac{6s(g-a)}{g}}w, x^{12s})}.\label{prop}
\end{equation}
\end{proposition}

Here the elliptic nome is $p=\e^{-\epsilon}$, $w = \e^{-2 \pi u / \epsilon}$,
and
$x = \e^{-\pi^2 /r \epsilon}$. Regime 2 is specified by the range of
  the spectral parameter: $0<u<3
\lambda$, and the value of
the crossing parameter: $\lambda=\pi s/r$ where $s=L+2$ and
$r=4(L+1)$.
For the dilute A$_4$ model the E$_7$ Coxeter number is $g=18$, while for the
A$_6$ model the E$_6$ Coxeter number is $g=12$.
The standard (conjugate modulus) elliptic function is defined by
\begin{equation}
E(z,q)=\prod_{n=1}^{\infty}(1-q^{n-1}z)(1-q^n/z)(1-q^n).
\nonumber
\end{equation}
The numbers  $a$ appearing in (\ref{prop}) are given in Tables \ref{eight},
\ref{seven} and
\ref{six}. The integers in these tables have appeared in other
contexts in relation to the E-algebras \cite{kos,bcds}.
%
%

In this paper we explicitly derive the elementary excitation spectrum of the
dilute A$_4$ model, thereby confirming our Proposition in this case.
The result (\ref{prop}) leads to the inverse correlation lengths and mass gaps.
Our input to these calculations are the string solutions to the Bethe
equations
found by Grimm and Nienhuis \cite{GNa,GN,GNp}.
As discussed later in \ref{disc}, our results are applicable to the tricritical
  Ising model
which is in the
same universality class.
In particular, the elliptic nome appearing in the dilute A$_4$ weights
in regime 2 corresponds to the
leading thermal off-critical perturbation in the
tricritical Ising model. This perturbation is
identified with $\phi_{(1,2)}$ \cite{lmc} and has been shown to
exhibit E$_7$ structures
\cite{FZ,cm}.
We are able to obtain exact results for some universal amplitudes
of the tricritical Ising model.
These results are in agreement with those found recently by other means
\cite{FMS,FMS2}.

The outline of the paper is as follows. The dilute A$_L$ lattice model
is defined along with the corresponding Bethe equations in \ref{model}.
The bulk free energy and the eigenvalue expressions in regime 2 for $L=4$
associated with the seven E$_7$ masses are derived via the exact perturbation
approach in \ref{spectrum}  (continued in the Appendix).
The paper concludes in \ref{disc} with a discussion of the results and their
relevance to universal behaviour in the tricritical Ising model.

\section{THE DILUTE A$_4$ MODEL}\label{model}

We here give a short summary of facts about the dilute A$_L$ models
\cite{WPSN, BS2} which are pertinent to our calculations.

The dilute A$_L$ model is an exactly solvable, $L$-state
restricted solid-on-solid
model defined on the square lattice. Its adjacency diagram is the Dynkin
diagram of A$_L$ with the additional possibility that a state may be adjacent
to itself on the lattice.
The model is solvable in four
off-critical regimes, with the elliptic nome $p$ of its Boltzmann weights
taking the
model off-critical. At criticality, the dilute A$_L$ model can be
constructed \cite{WNS,R} from
the dilute O($n$)
loop model \cite{N,WN}. In regime 2 of the model the central charge is
\[
c=1-\frac{6}{L(L+1)}.
\]

In the majority of exactly solved models the elliptic
nome plays the role of temperature \cite{Baxter}.
In the dilute A$_L$ model
the interpretation of the elliptic nome differs according
to whether $L$ is even or odd. For $L$ odd
the elliptic nome plays the role of a magnetic field \cite{WNS}, and $p>0$
and $p<0$ are related
by simple label reversal of the heights. For $L$ even the nome plays a
thermal role, and the
behaviour of the model depends on whether $p>0$ (regime 2$^+$) or $p<0$
(regime 2$^-$). More
specifically, it was shown \cite{WPSN} that in regime 2 the
nome corresponds to perturbation of the $\mathcal{M}_{L, L+1}$ minimal
unitary conformal field theories
by the operator $\phi_{(1,2)}$.

  Using the conjugate variables
introduced after (\ref{prop}), and setting $w_j = \e^{-2\pi
u_j/\epsilon}$, the eigenvalues of the row transfer matrix of the
dilute A models
(for a periodic strip of width $N$ where for convenience $N$ has been taken
as even) can be written \cite{BNW}
\begin{align}
\Lambda(w) =& \omega \left[
\frac{E(x^{4s}/w,x^{2r})\;E(x^{6s}/w,x^{2r})}
      {E(x^{4s},x^{2r})\;E(x^{6s},x^{2r})}
\right]^N
\prod_{j=1}^N w_j^{1 - 2s/r} \,
\frac{E(x^{2s} w /w_j,x^{2r})}{E(x^{2s} w_j/w,x^{2r})}
\nonumber \\
&\  +\, \left[
\frac{x^{2s}}{w}
\frac{E(w,x^{2r})\;E(x^{6s}/w,x^{2r})}
      {E(x^{4s},x^{2r})\;E(x^{6s},x^{2r})}
\right]^N
\nonumber \\
& \phantom{ w_j^{1 - 2s/r} \,\frac{E(x^{2s} w /w_j,x^{2r})}{E(x^{2s}
w_j/w,x^{2r})}}
\times
\prod_{j=1}^N w_j \,
\frac{E(w/w_j,x^{2r})\;E(x^{6s} w_j /w,x^{2r})}
      {E(x^{2s} w_j/w,x^{2r})\;E(x^{4s} w_j /w,x^{2r})}
\nonumber\\
&\  +\, \omega^{-1}
\left[ x^{2s} \,
\frac{E(w,x^{2r})\;E(x^{2s}/w,x^{2r})}
      {E(x^{4s},x^{2r})\;E(x^{6s},x^{2r})}
\right]^N \prod_{j=1}^N w_j^{2s/r}
\frac{E(x^{8s} w_j /w,x^{2r})}{E(x^{4s} w_j /w,x^{2r})}
\nonumber \\
\label{eigsc}
\end{align}
where $\omega=\exp(\text{i} \pi \ell/(L+1))$ for $\ell=1,\ldots,L$,
and $s=L+2$ and $r=4(L+1)$ in regime 2.
The Bethe equations which give the $N$ roots $u_j$
have the form
\begin{multline}
\omega \left[ w_j \,
\frac{E(x^{2s}/w_j,x^{2r})}{E(x^{2s} w_j,x^{2r})}
\right]^N \\
= - \prod_{k=1}^N w_k^{2s/r}
\frac{E(x^{2s} w_j/w_k,x^{2r}) \, E(x^{4s} w_k/w_j,x^{2r})}
      {E(x^{2s} w_k/w_j,x^{2r}) \, E(x^{4s} w_j/w_k,x^{2r})} .
\label{BAEc}
\end{multline}

In the limit $|p|\to 1$ with $u/\epsilon$ fixed, or equivalently $x \to 0$,
the excitations  in the eigenspectrum
$r_j(w)$, defined in (\ref{excite}),
break up into a number of distinct bands labelled by integer powers of $w$.
Numerical investigations of the eigenspectrum \cite{BNW, GNa, GN, GNp, BS3}
  have revealed eight and seven thermodynamically significant excitations
for $L=3$ and $L=4$
respectively, and provided the data in Table \ref{strings}.

We previously \cite{BS, BS2}  applied the
exact perturbation approach
initiated by  Baxter \cite{B72} to calculate the excitations in
the eigenspectrum  for
$L=3$. This involved perturbing away from the strong magnetic
field limit at $p\to 1$; for $L=4$ this limit corresponds to moving far away
from the critical
temperature. The calculations follow.

\section{MASS SPECTRUM}\label{spectrum}

\subsection{Preliminaries}

To apply the perturbation technique \cite{B72} to find the form of the
excitations (\ref{eigsc}),
the string structure of
the Bethe ansatz roots (\ref{BAEc}) is required input.
The groundstate roots all have
$u_j$ pure imaginary, so that
$w_j=\e^{-2\pi u_j/\epsilon}=a_j$ for $j=1, \ldots, N$ with $|a_j|=1$; in
this sense they all live on a
unit circle. For each excitation $i$, certain roots acquire a real part $m
\pi/20$, as shown in
Table \ref{strings}.  (If there are $n_i$ such roots, one says there is an
$n_i$-string
associated with the excitation.) For these roots
$w_j=b_j x^m$, so that the string entries can be thought of as living on
circles of radius $x^m$
with phase
$b_j$, while the other $N-n_i$ roots again lie on the unit circle.


The process of finding the excitations involves using the Bethe equations
(\ref{BAEc}) to set up recurrence relations
for auxiliary functions of the unknown roots $a_j$. As the roots only enter
the eigenvalue
expression (\ref{eigsc}) through the auxiliary functions, it just remains
to solve the
recurrence relations by iteration and to simplify the resulting
expressions. The largest
eigenvalue $\Lambda_0$, relative to which excitations are measured, was
calculated previously in this way \cite{BS2} for all $L$.

The relationship between the excitations (\ref{excite}), the correlation
lengths $\xi_j$ and
the mass spectrum $m_j$ of the associated field theory is
\begin{equation}
\xi_j^{-1}=-\log r_j =m_j, \label{rule}
\end{equation}
where we take the isotropic value $u=3 \lambda/2$.

It is convenient to use the notation for products:
\begin{align*}
(z;p_1, \ldots, p_k)_{\infty}&=\prod_{n_1, \ldots
n_k=0}^{\infty}(1-p_1^{n_1}\cdots
p_k^{n_k}z)\\
(z_1, \ldots z_m;p_1, \ldots, p_k)_{\infty}&=\prod_{j=1}^{m}(z_j;p_1, \ldots,
p_k)_{\infty}
\end{align*}
which satisfy many identities, the ones used repeatedly in what follows being:
\begin{align*}
\frac{(z;p)_{\infty}}{(zp;p)_{\infty}}&=(1-z)\\
\frac{(z;p,q)_{\infty}}{(zp;p,q)_{\infty}}&=(z;q)_{\infty}\\
\frac{(zq/p;p,q)_{\infty}}{(z;p,q)_{\infty}}&=
\frac{(zq/p;q)_{\infty}}{(z;p)_{\infty}}.
\end{align*}
The standard elliptic function is thus re-written as
\begin{equation}
E(z,q)=\prod_{n=1}^{\infty}(1-q^{n-1}z)(1-q^n/z)(1-q^n)=(z,q/z,q;q)_
{\infty}.\label{rewrite}
\end{equation}
It  also proves convenient to use the
shorthand notation $\prod_{j=1}^m a_j=A_m$.

For each $m_i$, if the associated string of excited roots has length $n_i$, we
define the required auxiliary functions of the as-yet-unknown roots to be
\begin{align}
F_i(w)&=\prod_{j=1}^{N-n_i} (w/a_j;x^{2r})_{\infty},
\nonumber\\
G_i(1/w)&=\prod_{j=1}^{N-n_i} (x^{2r}a_j/w;x^{2r})_{\infty}.
\label{defi}
\end{align}
In fact, we actually solve for combinations of these:
\begin{align}
\F_i(w)&=F_i(w)/F_i(x^{16}w)=F_i(w)/F_i(x^{2r-4s}w), \nonumber\\
\G_i(1/w)&=G_i(1/w)/G_i(1/x^{16}w)=G_i(1/w)/G_i(1/x^{2r-4s}w), \label{bigfg}
\end{align}
for $i =2,4,5,6,7$ (but for $i=1, 3$ slightly different definitions are
convenient and are
given as required).

So far as possible, we write factors and powers which are common to all
eigenvalues (or
indeed to all the eigenvalues for other A$_L$ models) in terms
of the generic $r$ and $s$ to distinguish them from the particular integers
which arise from the
input strings. Of course, $r=20$ and $s=6$ throughout. Once the particular
string form for the
roots has been applied, the calculations are straightforward for all masses
except $m_1$ and
$m_3$. For this reason, we sketch below the details for the first three masses.
The other cases follow
similar paths to $m_2$ or indeed to most of the masses for the dilute A$_3$
model \cite{BS2}, so we relegate them to the appendix. We
make some comments concerning
$m_1$,
$m_3$ and $m_6$ later on.


\subsection{Mass $m_1$}
We begin the perturbation argument with the structure
$w_j=a_j$ for $j=1, \ldots, N-3$ with $w_{N-1}=b_1x^{-4}$,
$w_{N-2}=b_2x^{4}$ and
$w_N=b_3x^{20}$, so that the string length is $n_1=3$. From the Bethe
equations (\ref{BAEc}) for
$j=N-2$,
$j=N-1$ and $j=N$ in the limit $x\to 0$ we can show that $b_1=b_2=b_3=b$.
The Bethe equation for the other roots $a_k=a$ is then
\begin{multline}
-\omega \left[a\frac{E(x^{2s}/a)}{E(x^{2s}a)}\right]^N=
(A_{N-3}b^3)^{3/5}\frac{a^2}{b^2}
\\
\times\frac{E(x^{4} b/a)E(x^{24} b/a)E(x^{28} b/a)}{E(x^{4} a/b)E(x^{24}
a/b)E(x^{28} a/b)}
\prod_{j=1}^{N-3} \frac{E(x^{2s}a/a_j)E(x^{4s}a_j/a)}
{E(x^{2s}a_j/a)E(x^{4s}a/a_j)}.\label{bethe1.k}
\end{multline}
In the $x \to 0$ limit this gives the equation
\begin{equation}
a^{N-2}+\frac{1}{\omega} (A_{N-3}b^3)^{3/5}/b^2=0, \label{aeq}
\end{equation}
which is an equation of order $(N-2)$, so that there is a missing root on
the unit
circle, a `hole', which we call $a_{N-2}$.  Since this is an equation for
the roots, its left
hand side must be equivalent to
$\prod_{j=1}^{N-2}(a-a_j)$, and equating the constant terms from these two
expressions we obtain
\begin{equation}
  \frac{1}{\omega} (A_{N-3}b^3)^{3/5}=A_{N-2}b^2=A_{N-3}a_{N-2}b^2,
\label{coeff}
\end{equation}
(which we later apply to prefactors in $\Lambda_1$).
The Bethe equations for $b$ taken together in this limit, and combined with
(\ref{coeff}) give
\begin{equation*}
\left[\frac{1}{\omega} (A_{N-3}b^3)^{3/5}\right]^3=-b^6 (A_{N-3})^2 \quad
\Rightarrow \quad
A_{N-3}(a_{N-2})^3=-1 .
\end{equation*}
We use this, together with the fact that each root $a_j$, including the
hole, must satisfy
(\ref{aeq}), to show
\begin{equation}
(a_{N-2})^{N-2}=-A_{N-3}a_{N-2} \quad \Rightarrow \quad (a_{N-2})^{N}=1.
\label{apower}
\end{equation}

We define the following auxiliary functions of the roots (see (\ref{defi})):
\begin{align*}
\F_1(w)&=\frac{F_1(w)}{F_1(x^{16}w)}\frac{(x^4w/b;x^{2r})_{\infty}}{(x^{12}w/b
;x^{2r})_{\infty}},\\
\G_1(1/w)&=\frac{G_1(1/w)}{G_1(1/x^{16}w)}
\frac{(x^{36}b/w;x^{2r})_{\infty}}{(x^{24}b/w;x^{2r})_{\infty}}.
\end{align*}
They must satisfy  recurrence relations arising from (\ref{bethe1.k})
\begin{align}
\F_1(a)&=\left[\frac{(x^{2s}a;x^{2r})_{\infty}}{(x^{2r-2s}a;x^{2r})_{\infty}}
\right]^N
\frac{(x^{24}a/a_{N-2},x^{28}a/a_{N-2};x^{2r})_{\infty}}
{(x^{12}a/a_{N-2},x^{16}a/a_{N-2};x^{2r})_{\infty}}
\frac{\F_1(x^{2s}a)}{\F_1(x^{4s}a)}, \nonumber \\
\G_1(1/a)&=\left[\frac{(x^{2r+2s}/a;x^{2r})_{\infty}}{(x^{6s}/a;x^{2r})_
{\infty}}\right]^N
\frac{(x^{36}a_{N-2}/a,x^{40}a_{N-2}/a;x^{2r})_{\infty}}
{(x^{48}a_{N-2}/a,x^{52}a_{N-2}/a;x^{2r})_{\infty}}
\frac{\G_1(x^{2s}/a)}{\G_1(x^{4s}/a)} .\label{recc}
\end{align}
Solving these we obtain
\begin{align}
\F_1(a)&=\F_0(a)
\frac{(x^{40}a/a_{N-2};x^{2r})_{\infty}}{(x^{16}a/a_{N-2};x^{2r})_{\infty}}
\frac{(x^{36}a/a_{N-2},x^{48}a/a_{N-2};x^{12s})_{\infty}}
{(x^{12}a/a_{N-2},x^{72}a/a_{N-2};x^{12s})_{\infty}}, \nonumber
\\
\G_1(1/a)&=\G_0(1/a)
\frac{(x^{40}a_{N-2}/a;x^{2r})_{\infty}}{(x^{64}a_{N-2}/a;x^{2r})_{\infty}}
\frac{(x^{36}a_{N-2}/a,x^{96}a_{N-2}/a;x^{12s})_{\infty}}
{(x^{60}a_{N-2}/a,x^{72}a_{N-2}/a;x^{12s})_{\infty}}. \label{sol1}
\end{align}
Here $\F_0$ and $\G_0$ arise from the square bracketed factors in
(\ref{recc}) and give rise to
the square bracketed factor in (\ref{first}). They are related to the
groundstate eigenvalue
$\Lambda_0$, they are common to the calculation of each mass and we will
suppress these factors
for
$m_2,
\ldots, m_7$. We now write the eigenvalue expression in terms of the
auxiliary functions, the
first term
being
\begin{multline}
\frac{\Lambda_1}{3}=-\frac{w}{a_{N-2}}\left[\frac{(x^{2r-6s}w,x^{2r-4s 
}w,x^{4s}/
w,x^{6s}/w;x^{2r})_{\infty}}
{(x^{2r-6s},x^{2r-4s},x^{4s},x^{6s};x^{2r})_{\infty}}\right]^N \\
\times
\frac{(x^{28}w/a_{N-2},x^{12}a_{N-2}/w;x^{2r})_{\infty}}{(x^{12}w/a_{N-2},
x^{28}a_{N-2}/w;x^{2r})_{\infty}}
\F_1(x^{2s}w)\G_1(1/x^{2s}w). \label{first}
\end{multline}
Substituting the solutions (\ref{sol1}) gives an expression for the
excitation $r_1(w)$ which may be
written in elliptic
functions (\ref{rewrite}) as
\begin{equation}
\frac{\Lambda_1}{\Lambda_0} = w\,
\frac{E(-x^{12}/w,x^{12s})E(-x^{48}w,x^{12s})}
      {E(-x^{12}w,x^{12s})E(-x^{48}/w,x^{12s})},\label{eig1}
\end{equation}
where we have set
$a_{N-2}=-1$.
(The other two terms in the eigenvalue always give identical elliptic
function expressions
to the first, upon simplification.)

The Bethe equations involving $b$ and the `hole' equation, which is
(\ref{bethe1.k}) with
$a=a_{N-2}$, can also be expressed in terms of the auxiliary functions.
Application of
identities and simplification gives:
\begin{gather*}
E(x^{12} b/a_{N-2}, x^{2r-4s})=E(x^{12} a_{N-2}/b, x^{2r-4s})\\
\left[\frac{E(x^{12}a_{N-2}, x^{12s})E(x^{48}/a_{N-2}, x^{12s})}
{E(x^{12}/a_{N-2}, x^{12s})E(x^{48}a_{N-2}, x^{12s})}\right]^N=(a_{N-2})^N.
\end{gather*}
Clearly $a_{N-2}=b=-1$ (identified initially from numerical studies)
satisfy these conditions;
the second reduces to (\ref{apower}) in the $x \to 0$ limit, and note the
similarities with (\ref{eig1}).

\subsection{Mass $m_2$}
We begin the perturbation argument with the structure
$w_j=a_j$ for $j=1, \ldots, N-2$ with $w_{N-1}=b_1x^{-14}$ and
$w_N=b_2x^{14}$, so that $n_2=2$. From the Bethe equations for $j=N-1$ and
$j=N$
we can show that $b_1=b_2=b$.
The Bethe equation for the other roots $a_k=a$ is then
\begin{multline}
-\omega \left[a\frac{E(x^{2s}/a)}{E(x^{2s}a)}\right]^N=
(A_{N-2}b^2)^{3/5}\frac{a^2}{b^2}
\\
\times \frac{E(x^{10} b/a)E(x^{14} b/a)}{E(x^{10} a/b)E(x^{14} a/b)}
\prod_{j=1}^{N-2} \frac{E(x^{2s}a/a_j)E(x^{4s}a_j/a)}
{E(x^{2s}a_j/a)E(x^{4s}a/a_j)}.\label{bethe2.k}
\end{multline}
In the $x \to 0$ limit this gives the equation
\begin{equation*}
a^{N-2}+\frac{1}{\omega} (A_{N-2}b^2)^{3/5}/b^2=0,
\end{equation*}
which has the same order as the number of unknown roots ($N-2$) so that
there is no hole.
Equating this with
$\prod_{j=1}^{N-2}(a-a_j)$ we obtain
\begin{equation*}
  \frac{1}{\omega} (A_{N-2}b^2)^{3/5}=A_{N-2}b^2
\end{equation*}
(which we later apply to prefactors in $\Lambda_2$).
{}From the other Bethe equations
in this limit,
\begin{equation}
\left[\frac{1}{\omega} (A_{N-2}b^2)^{3/5}\right]^2=\frac{( A_{N-2}b^2)^2}
{b^{2N}}\quad \Rightarrow \quad b^{2N}=1.\label{bpower}
\end{equation}

Treating the Bethe equation (\ref{bethe2.k}) as before gives, in terms
of the functions defined in (\ref{defi}) and (\ref{bigfg}), the recurrences
\begin{align*}
\F_2(a)&=
\frac{(x^{26}a/b,x^{30}a/b;x^{2r})_{\infty}}{(x^{10}a/b,x^{14}a/b;x^{2r})_
{\infty}}
\frac{\F_2(x^{2s}a)}{\F_2(x^{4s}a)}, \nonumber \\
\G_2(1/a)&=
\frac{(x^{38}b/a,x^{34}b/a;x^{2r})_{\infty}}{(x^{50}b/a,x^{54}b/a;x^{2r})_
{\infty}}
\frac{\G_2(x^{2s}/a)}{\G_2(x^{4s}/a)} .
\end{align*}
Solving these we obtain
\begin{align*}
\F_2(a)&=
\frac{(x^{30}a/b,x^{42}a/b;x^{2r})_{\infty}}{(x^{14}a/b,x^{26}a/b;x^{2r})_
{\infty}}
\frac{(x^{26}a/b,x^{38}a/b,x^{46}a/b,x^{58}a/b;x^{12s})_{\infty}}
{(x^{10}a/b,x^{22}a/b,x^{62}a/b,x^{74}a/b;x^{12s})_{\infty}},\\
\G_2(1/a)&=
\frac{(x^{38}b/a,x^{50}b/a;x^{2r})_{\infty}}{(x^{54}b/a,x^{66}b/a;x^{2r})_
{\infty}}
\frac{(x^{34}b/a,x^{46}b/a,x^{86}b/a,x^{98}b/a;x^{12s})_{\infty}}
{(x^{50}b/a,x^{62}b/a,x^{70}b/a,x^{82}b/a;x^{12s})_{\infty}}.
\end{align*}
We now substitute these into the eigenvalue expression,
the first term of which is
\begin{equation*}
\frac{\Lambda_2}{3}=\frac{w^2}{b^2}
  \frac{(x^{26}w/b,x^{38}w/b,x^{2}b/w,x^{14}b/w;x^{2r})_{\infty}}{(x^{2 
}w/b,x^{14
}w/b,x^{26}b/w,
x^{38}b/w;x^{2r})_{\infty}}
\F_2(x^{2s}w)\G_2(1/x^{2s}w).
\end{equation*}
This gives an expression for the excitation in elliptic functions (setting
$b=-1$):
\begin{equation}
\frac{\Lambda_2}{\Lambda_0} = w^2\,
\frac{E(-x^{2}/w,x^{12s})\,E(-x^{14}/w,x^{12s})
E(-x^{38}\,w,x^{12s})\,E(-x^{50}
w,x^{12s})}
{E(-x^{2}\,w,x^{12s})\,E(-x^{14}\,w,x^{12s})E(-x^{38}/w,x^{12s})\,E(-x 
^{50}/w,x^
{12s})}.\label{eig2}
\end{equation}

If the product of the six Bethe equations involving $b$ is expressed in
terms of the auxiliary
functions, the equation for $b$ (generalizing $b^{2N}=1$ seen in the $x \to
0$ limit in
(\ref{bpower})) is clearly satisfied by $b=-1$:
\begin{equation*}
\left[\frac{E(x^{2}b,x^{12s})\,E(x^{14}b,x^{12s})E(x^{38}/b,x^{12s})\,E(x^{5
0}/b,x^{12s})}    {E(x
^{2}/b,x^{12s})\,E(x^{14}/b,x^{12s})E(x^{38}b,x^{12s})\,E(x^{50}b,x^{12s})}
\right]^N=b^{2N}.
\end{equation*}
Compare the pattern of powers of $x$ in this equation with those in
(\ref{eig2}); this equation
has a precise analogue for each mass $m_4, \ldots, m_7$, which will not be
given.


\subsection{Mass $m_3$}
We begin the perturbation argument with the string structure
$w_j=a_j$ for $j=1, \ldots, N-3$ with $w_{N-2}=b_1x^{-12}$,
$w_{N-1}=b_2x^{12}$ and $w_{N}=b_3x^{20}$. From the Bethe equations for
$j=N-2$ and $j=N-1$
we can show that $b_1=b_2=\alpha$, but the Bethe equation for $j=N$ does not
link $b_3=b$ to $\alpha$ in the $x\to 0$ limit. (This feature was observed
also in the $L=3$
case, for a string of odd length \cite{BS2}.)
The Bethe equation for the other roots $a_k=a$ is then
\begin{multline}
-\omega
\left[a\frac{E(x^{2s}/a)}{E(x^{2s}a)}\right]^N=(A_{N-3}\alpha^2b)^{3/5}
\frac{a^3}{\alpha b^2}
\frac{E(x^{4} b/a)E(x^{8} b/a)}{E(x^{4} a/b)E(x^{8} a/b)}
\\
\times \frac{E(x^{12} \alpha /a)E(x^{16} \alpha /a)E(x^{36}\alpha /
a)}{E(x^{12}
a/\alpha)E(x^{16} a/
\alpha)E(x^{36} a/\alpha)}
\prod_{j=1}^{N-3} \frac{E(x^{2s}a/a_j)E(x^{4s}a_j/a)}
{E(x^{2s}a_j/a)E(x^{4s}a/a_j)}.\label{bethe3.k}
\end{multline}
In the $x \to 0$ limit this gives the equation
\begin{equation*}
a^{N-3}-\frac{1}{\omega} (A_{N-3}\alpha^2 b)^{3/5}/ \alpha b^2=0.
\end{equation*}
Equating this as usual with
$\prod_{j=1}^{N-3}(a-a_j)$, we obtain
\begin{equation*}
  \frac{1}{\omega} ((A_{N-3}\alpha^2 b)^{3/5}=A_{N-3}\alpha b^2
\end{equation*}
(which we later apply to prefactors in $\Lambda_3$).
{}From the other Bethe equations
in this limit,
\begin{equation*}
\left[\frac{1}{\omega} (A_{N-3}\alpha^2 b)^{3/5}\right]^3=\frac{(
A_{N-3}\alpha b^2)^3}
{b^{2N}}\quad \Rightarrow \quad b^{2N}=1.
\end{equation*}

In this case it is convenient to define
\begin{align}
\F_3(w)&=\frac{F_3(w)}{F_3(x^{16}w)}\frac{(x^{12}w/ \alpha
;x^{2r})}{(x^{4}w/ \alpha ;x^{2r})},
\nonumber\\
\G_3(1/w)&=\frac{G_3(1/w)}{G_3(1/x^{16}w)}\frac{(x^{28}\alpha /w
;x^{2r})}{(x^{36} \alpha /w
;x^{2r})}, \label{def3}
\end{align}
because this choice will make it clear that $\alpha$ is a spectator in the
solution to the
recurrence relation; it does not appear in the eigenvalue expression.

Treating the Bethe equation (\ref{bethe3.k}) as before gives the recurrences
\begin{align*}
\F_3(a)&=
\frac{(x^{32}a/b,x^{36}a/b;x^{2r})_{\infty}}{(x^{4}a/b,x^{8}a/b;x^{2r} 
)_{\infty}
}
\frac{\F_3(x^{2s}a)}{\F_3(x^{4s}a)}, \nonumber \\
\G_3(1/a)&=
\frac{(x^{56}b/a,x^{60}b/a;x^{2r})_{\infty}}{(x^{28}b/a,x^{32}b/a;x^{2 
r})_{\infty}}
\frac{\G_3(x^{2s}/a)}{\G_3(x^{4s}/a)} .
\end{align*}
Solving these we obtain
\begin{align*}
\F_3(a)&=
\frac{(x^{36}a/b;x^{2r})_{\infty}}{(x^{20}a/b;x^{2r})_{\infty}}
\frac{(x^{32}a/b,x^{40}a/b,x^{44}a/b,x^{52}a/b;x^{12s})_{\infty}}
{(x^{4}a/b,x^{8}a/b,x^{16}a/b,x^{68}a/b;x^{12s})_{\infty}},
\\
\G_3(1/a)&=
\frac{(x^{44}b/a;x^{2r})_{\infty}}{(x^{60}b/a;x^{2r})_{\infty}}
\frac{(x^{28}b/a,x^{32}b/a,x^{40}b/a,x^{92}b/a;x^{12s})_{\infty}}
{(x^{56}b/a,x^{64}b/a,x^{68}b/a,x^{76}b/a;x^{12s})_{\infty}}.
\end{align*}
We now substitute these into the eigenvalue expression,
the first term of which is, in terms of the functions (\ref{def3}),
\begin{equation*}
\frac{\Lambda_3}{3}=\frac{w^2}{b^2}
\frac{(x^{32}w/b,x^{8}b/w;x^{2r})_{\infty}}{(x^{8}w/b,x^{32}b/w;x^{2r} 
)_{\infty}
}
\F_3(x^{2s}w)\G_3(1/x^{2s}w).
\end{equation*}
With $b=-1$ this gives the expression in elliptic functions
\begin{equation}
\frac{\Lambda_3}{\Lambda_0} = w^2\,
\frac{E(-x^{8}/w,x^{12s})\,E(-x^{16}/w,x^{12s})E(-x^{44}\,w,x^{12s})\, 
E(-x^{52}\
,w,x^{12s})}
{E(-x^{8}\,w,x^{12s})\,E(-x^{16}\,w,x^{12s})E(-x^{44}/w,x^{12s})\,E(-x 
^{52}/w,x^
{12s})}.\label{eig3}
\end{equation}
The Bethe equations involving $\alpha$ and $b$, also expressed in terms of
the auxiliary
functions, give:
\begin{gather*}
E(x^{12} b/\alpha, x^{2r-4s})=E(x^{12} \alpha/b, x^{2r-4s})\\
\left[\frac{E(x^{8}b, x^{12s})E(x^{16}b, x^{12s})E(x^{44}/b,
x^{12s})E(x^{52}/b, x^{12s})}{E(x^{8}/b, x^{12s})E(x^{16}/b,
x^{12s})E(x^{44}b, x^{12s}
)E(x^{52}b,
x^{12s})}\right]^N=b^{2N}.
\end{gather*}
Notice that $b=-1$ satisfies this second equation, and the (so far missing)
link between $\alpha$
and $b$ is provided by the first.

\subsection{Comments on the `odd' strings}

To close this rather technical section of the paper, we wish to briefly
comment on
the strings of odd length (see Table \ref{strings}).

For this model strings of odd length appear for the first, third and sixth
masses. In the first case, the odd string of excited roots is accompanied
by a `hole' among the roots
on the unit circle; it is only this hole $a_{N-2}$ which appears in the
eigenvalue expression. In the case of
the third mass, the phase $b$ of the string entry $m=10$ appears alone in
the eigenvalue expression, and the
other entries of the string seem to have a spectator role. (For the sixth
mass, there was nothing
special about the calculation.) In the calculations for dilute A$_3$
\cite{BS2} odd
strings were involved for masses 4 and
6, where again the calculation of the associated excitation was less
straightforward than for even strings. In
one case, both the coefficient of the `odd' entry {\em and} a hole appeared
in the eigenvalue, while the other
calculation resembles that of $m_3$ in this paper. In general the string
entries come
in pairs $\pm m$, except for $m=r/2$, which stands alone if it occurs, due
to the period of the original
elliptic functions in $u_j$. This is the only source of strings of odd
length; we can only conclude that when
such an entry occurs, it in some sense dominates
calculations following the
exact perturbation technique.
For strings of even length, all the excited roots seem to contribute in a
more equal fashion to the
calculation and to the resulting eigenvalue expression.

\section{DISCUSSION}\label{disc}

In this paper we have made use of the Bethe Ansatz string solutions found
by Grimm and
Nienhuis to derive the excitation spectrum of the dilute A$_4$ model via the
exact perturbation approach. Our expressions for the seven
thermodynamically significant excitations for
the dilute
A$_4$ eigenspectrum in regime 2$^-$ are given in
(\ref{eig1}), (\ref{eig2}), (\ref{eig3}), (\ref{eig4}), (\ref{eig5}),
(\ref{eig6}) and
(\ref{eig7}).
In this way we have verified for a second case the
Proposition  given by (\ref{prop}).

It is perhaps unsatisfying that an elegant closed form
expression such as (\ref{prop}) has been confirmed in the
A$_4$ case by
relying on numerical data for the strings (Table
\ref{strings}). Indeed, as described for the A$_3$ case in
\cite{BS2},  and in the detailed study 
\cite{GN}, tracing the strings from
$p=0$ (criticality) to the position they take in the scaling
(massive) limit reveals complicated structure 
(reported with one difference by two
groups of authors \cite{BNW, GN}). 
Fortunately, (\ref{prop}) was conjectured
\cite{BS3, SB} on the basis of general properties of the dilute
A models and of the E-type algebras, known to be linked by
their common connection to the $\phi_{(1,2)}$ perturbation
of the minimal unitary series; the (scaling limit) string data
used here has not contradicted it, and (admittedly limited)
numerical studies agreed with the lower eigenvalues \cite{BS3}.
A forthcoming paper
\cite{KS} should shed some new light, from the perspective of
Coxeter geometry, on the excitations (\ref{prop}) and hence,
among other things, on the string conjectures to which they are
related as demonstrated here in the $L=4$ case.

Recall that the central charge for dilute A$_4$ in regime 2 is
$c=\frac{7}{10}$.
There are several other known manifestations of the $c = \frac{7}{10}$ theory.
The Blume-Capel model \cite{C} is related to the Blume-Emery-Griffiths
model \cite{BEG}, a classical spin-1 Ising model with lattice
Hamiltonian
\begin{equation}
\mathcal{H}_{\rm BEG} = - J \sum_{\langle i,j \rangle} S_i S_j - D 
\sum_i (1-S_i^2)
- H \sum_i S_i-H_3\sum_{\langle i,j \rangle}S_iS_j(S_i+S_j), \label{ham}
\end{equation}
where $J$ is the nearest-neighbour interaction,
$D$ is a crystal field, $H$ a magnetic field term and $H_3$ is a
staggered magnetic field. The phase diagrams of these models
exhibit a tricritical point, as had been observed in
physical systems \cite{LS}.

The critical exponents, known from renormalization
group studies, are related to the Kac table of the
$c=\frac{7}{10}$ conformal field theory \cite{Hbook}.

After the Ising critical point, the universality class of the
tricritical Ising model corresponds to the
second simplest unitary conformal field theory
in two dimensions.
It is also the first of the super-conformal minimal models.
It can be perturbed by its four relevant scaling fields, shown in
Table \ref{TIM} ordered according to the associated conformal weight.
The leading magnetic perturbation is believed to be non-integrable
\cite{lmc}, and each of the other three perturbations give integrable quantum
field theories.
In the scaling limit these can each be associated with a
solvable interaction round a face (IRF) model (or to the terms
in (\ref{ham})). The ABF A$_4$ model in regime III
\cite{ABF,hu}
realizes the subleading
thermal perturbation. A lattice realization of the subleading
magnetic perturbation is given by
the dilute A$_3$ model in regime 1 \cite{kas}, and the scaling
limit of
the leading thermal perturbation corresponds to the
dilute A$_4$ model as considered in this paper.


The leading thermal perturbation
is known to be integrable and massive,
the masses being described by E$_7$ Toda
field theory \cite{FZ,cm}.
Numerical results from a finite-size analysis in the
spin-chain formulation \cite{g}, and from field theory via the truncated
conformal space approach
\cite{lmc} demonstrated the first few masses.

These are:
\begin{equation}
\begin{array}{lll}
m_1 = 1 & & {\rm ~odd} \\
m_2 = 2 \cos \frac{5\pi}{18}  & = 1.285~575\ldots & {\rm ~even}\\
m_3 = 2 \cos \frac{\pi}{9} & = 1.879~385\ldots & {\rm ~odd}\\
m_4 = 2 \cos \frac{\pi}{18} & = 1.969~615\ldots & {\rm ~even}\\
m_5 = 4 \cos \frac{\pi}{18} \cos \frac{5\pi}{18} & = 2.532~088\ldots & {\rm
~even}\\
m_6 = 4 \cos \frac{\pi}{9} \cos \frac{2\pi}{9}  & = 2.879~385\ldots & {\rm
~odd}\\
m_7 = 4 \cos \frac{\pi}{18} \cos \frac{\pi}{9} & = 3.701~666\ldots & {\rm
~even}\\
\end{array}
\label{masses7}
\end{equation}

The mass spectrum can be
classified \cite{lmc}  into even and odd states
(as indicated in (\ref{masses7})) corresponding to the
${\mathbb Z}_2$ symmetry of the affine E$_7$ Dynkin diagram.
Each of the above seven masses appears in the high-temperature
phase of the tricritical Ising model.
However, only the even subset appears in the low-temperature phase.
This is consistent with the numerical observations on the eigenspectrum of the
dilute A$_4$ model \cite{GNp,BS3}. For regime 2$^+$, in a study of 
the low-lying
excitations, the first and third were absent.
As we have demonstrated, all seven excitations are present in regime 
2$^-$ which
(through a quirk in labelling) corresponds to $T>T_c$.

Our expression (\ref{prop}) gives the correlation lengths and related masses
(\ref{rule}), expressed in terms of standard elliptic functions and the
original nome $p$, as
\begin{equation}
m_j=\xi_j^{-1}=2\sum_{a}
\log \frac{
\tv(\frac{a\pi}{36}+\frac{\pi}{4},p^{5/9})}
{\tv(\frac{a\pi}{36}-\frac{\pi}{4},p^{5/9})}.
\end{equation}
In the critical limit $p \to 0$ the leading order behaviour is
\begin{equation}
m_j \sim 8\, p^{5/9}  \sum_a \sin \tfrac{a\pi}{18}.\label{7masses}
\end{equation}
Substituting the integers of Table \ref{seven}, applying 
trigonometric identities
and taking mass ratios it was demonstrated \cite{BS3} that the E$_7$ mass
spectrum (\ref{masses7}) is recovered.

The ground states of the tricritical Ising model (in zero magnetic field)
have been identified \cite{C,lmc}.
For $T < T_c$, the system is in a
two-phase region of spontaneously broken spin reversal symmetry, with
two degenerate ground states in the thermodynamic limit. For $T > T_c$
there is one ground state.
This ground state picture is also consistent with that of the dilute A$_4$
model \cite{gs} as $|p| \to 1$.
In regime $2^+$ there are two possible ferromagnetic ground states,
while in regime $2^-$ there is a single disordered ground state.
(It is the presence of such disordered states
for $L$ even which complicates the calculation of order parameters for
this half of the dilute A$_L$ hierarchy.)

Very recently, an array of universal
ratios for the critical amplitudes of the tricritical Ising model have been
calculated \cite{FMS,FMS2} by field theoretic methods.
Not all of these quantities appear to be accessible via the
dilute A$_4$ model.
However, one such ratio involves the correlation length
prefactors $\xi_0^\pm$, above and below the critical temperature.
Our results and observations on the eigenspectrum of dilute A$_4$ give
this same value:
\begin{equation*}
\frac{\xi_0^+}{\xi_0^-} = \frac{\xi_1}{\xi_2} = 2 \cos \frac{5\pi}{18}.
\end{equation*}

We previously \cite{BS3} derived the amplitude
\begin{equation}
f_{\rm s} \xi_1^2 = \frac{1}{8 \sqrt{3}\cos(2 \pi/9)}=0.09420\ldots ,
\end{equation}
where $f_{\rm s}$ is the singular part of the free energy.
This agrees with the determination of this quantity for the $\phi_{(1,2)}$
perturbation of
the $c=\frac{7}{10}$ field theory \cite{F}. A related universal quantity is
the amplitude ratio associated with the correlation length \cite{PHA}
\begin{equation*}
R_{\xi}^{\pm}=A^{\frac{1}{2}} \xi_0^{\pm},
\end{equation*}
where $A/\alpha$ is the amplitude of the specific heat and $\alpha$ is
the related critical exponent. Our expressions for these quantities are
\begin{align*}
R_{\xi}^{+}&=\left[ \frac{10}{9^3\sqrt{3}\cos(2 \pi/9)} \right]^{\frac{1}{2}}
= 0.101678 \ldots\\
&\\
R_{\xi}^{-}&=\left[\frac{5}{2^3 9^2\sqrt{3}\cos(5\pi/18)\sin(5 
\pi/9)}\right]^{\frac{1}{2}}
= 0.083889 \ldots
\end{align*}
which agree with the numerical values of \cite{FMS2} (allowing for a
difference in definition by a factor $\alpha^{1/2}$ ).
As remarked \cite{FMS,FMS2}, such values may be observed in
experimental systems within the tricritical Ising universality class.

\section*{ACKNOWLEDGMENTS}
We thank Bernard Nienhuis and Uwe Grimm for making available to us their
unpublished results. We
also acknowledge interaction on this topic with Vladimir Bazhanov, and thank
Giuseppe Mussardo for his
interest in our work.
This paper was completed while
KAS enjoyed the hospitality of the University of Melbourne and
SUNY Stony Brook.

This work has been supported by the Australian Research Council.

\newpage
\renewcommand{\thesection}{APPENDIX \Alph{section}:}
\renewcommand{\theequation}{\Alph{section}\arabic{equation}}
\setcounter{section}{0}
\setcounter{equation}{0}

\appendix
\renewcommand{\thesection}{Appendix \Alph{section}}
\renewcommand{\thesubsection}{\arabic{subsection}}

\section{Further mass calculations}
\subsection{Mass $m_4$}
We begin the perturbation argument with the structure
$w_j=a_j$ for $j=1, \ldots, N-4$ with $w_{N-3}=b_1x^{-18}$,
$w_{N-2}=b_2x^{18}$, $w_{N-1}=b_3x^{-6}$ and
$w_N=b_4x^{6}$. From the Bethe equations for $j=N-3,\ldots,N$ we can show that
$b_1=b_2=b_3=b_4=b$.  The Bethe equation for the other roots is
\begin{multline}
-\omega \left[a\frac{E(x^{2s}/a)}{E(x^{2s}a)}\right]^N=
  \\
(A_{N-4}b^4)^{3/5}\frac{a^4}{b^4}\frac{E(x^2 b/a)E(x^{6} b/a)}{E(x^2
a/b)E(x^{6} a/b)}
\prod_{j=1}^{N-4} \frac{E(x^{2s}a/a_j)E(x^{4s}a_j/a)}
{E(x^{2s}a_j/a)E(x^{4s}a/a_j)}.\label{bethe4.k}
\end{multline}
In the $x \to 0$ limit this gives the equation
\begin{equation*}
a^{N-4}+\frac{1}{\omega} (A_{N-4}b^4)^{3/5}/b^4=0,
\end{equation*}
so that as usual we find expression involving the prefactors
\begin{equation*}
  \frac{1}{\omega} (A_{N-4}b^4)^{3/5}=A_{N-4}b^4.
\end{equation*}
Using this with the other Bethe equations
in the $x\rightarrow 0$ limit we obtain
\begin{equation*}
\left[\frac{1}{\omega} (A_{N-4}b^4)^{3/5}\right]^4=\frac{( A_{N-4}b^4)^4}
{b^{4N}}\quad \Rightarrow \quad b^{4N}=1.
\end{equation*}

From (\ref{bethe4.k}) come the recurrences
\begin{align*}
\F_4(a)&=
\frac{(x^{34}a/b,x^{38}a/b;x^{2r})_{\infty}}{(x^{2}a)/b,x^{6}a/b;x^{2r})_
{\infty}}
\frac{\F_4(x^{2s}a)}{\F_4(x^{4s}a)}, \nonumber \\
\G_4(1/a)&=
\frac{(x^{30}b/a,x^{26}b/a;x^{2r})_{\infty}}{(x^{58}b/a,x^{62}b/a;x^{2r})
_{\infty}}
\frac{\G_4(x^{2s}/a)}{\G_4(x^{4s}/a)}.
\end{align*}
The solutions are
\begin{align*}
\F_4(a)=& \frac{(x^{38}a/b,x^{42}a/b,x^{50}a/b,x^{54}a/b;x^{2r})_{\infty}}
{(x^{2}a/b,x^{6}a/b,x^{14}a/b,x^{18}a/b;x^{2r})_{\infty}}
\nonumber \\
&\times
\frac{(x^{34}a/b,x^{38}a/b,x^{46}a/b,x^{50}a/b;x^{12s})_{\infty}}
{(x^{70}a/b,x^{74}a/b,x^{82}a/b,x^{86}a/b;x^{12s})_{\infty}},
\\
\G_4(1/a)=&
\frac{(x^{26}b/a,x^{30}b/a,x^{38}b/a,x^{42}b/a;x^{2r})_{\infty}}
{(x^{62}b/a,x^{66}b/a,x^{74}b/a,x^{78}b/a;x^{2r})_{\infty}}
\nonumber \\
&\times
\frac{(x^{94}b/a,x^{98}b/a,x^{106}b/a,x^{110}b/a;x^{12s})_{\infty}}
{(x^{58}b/a,x^{62}b/a,x^{70}b/a,x^{74}b/a;x^{12s})_{\infty}}.
\end{align*}
In terms of these functions the eigenvalue may be represented as
\begin{equation*}
\frac{\Lambda_4}{3}=\frac{w^2}{b^2}
\frac{(x^{18}w/b,x^{30}w/b,x^{10}b/w,x^{22}b/w;x^{2r})_{\infty}}
{(x^{10}w/b,x^{22}w/b,x^{30}b/w,x^{18}b/w;x^{2r})_{\infty}}
\F_4(x^{2s}w)\G_4(1/x^{2s}w).
\end{equation*}
Thus, application of the perturbation argument yields the
excitation to be
\begin{equation}
\frac{\Lambda_4}{\Lambda_0}=w^2\,
\frac{E(-x^{10}/w,x^{12s})\,E(-x^{14}/w,x^{12s})E(-x^{46}w,x^{12s})\,E 
(-x^{50}w,
x^{12s})}
{E(-x^{10}w,x^{12s})\,E(-x^{14}w,x^{12s})E(-x^{46}/w,x^{12s})\,E(-x^{5 
0}/w,x^{12
s})},\label{eig4}
\end{equation}
where we have put $b=-1$.
\subsection{Mass $m_5$}
We begin the perturbation argument with
$w_j=a_j$ for $j=1, \ldots, N-4$ and  $w_{N-3}=b_1x^{-16}$,
$w_{N-2}=b_2x^{16}$,
$w_{N-1}=b_3x^{-12}$, $w_N=b_4x^{12}$. We can
show that the $b_i$ are equal, and we call them $b$.
The Bethe equation for the other roots is
\begin{multline}
-\omega \left[a\frac{E(x^{2s}/a)}{E(x^{2s}a)}\right]^N=
(A_{N-4}b^4)^{3/5}\frac{a^4}{b^4}
\frac{E(x^{8}b/a)E^2(x^{12}b/a)}{E(x^8a/b)E^2(x^{12} a/b)} \\
\times\frac{E(x^{16}b/a)}
{E(x^{16}a/b)}
\prod_{j=1}^{N-4} \frac{E(x^{2s}a/a_j)E(x^{4s}a_j/a)}
{E(x^{2s}a_j/a)E(x^{4s}a/a_j)}. \label{bethe5.k}
\end{multline}
In the $x \to 0$ limit this gives the equation
\begin{equation*}
a^{N-4}+\frac{1}{\omega} (A_{N-4}b^4)^{3/5}/b^4=0,
\end{equation*}
which leads in the usual way to a prefactor expression
\begin{equation*}
  \frac{1}{\omega} (A_{N-4}b^4)^{3/5}=A_{N-4}b^4.
\end{equation*}
From this and the other Bethe equations
\begin{equation*}
\left[\frac{1}{\omega} (A_{N-4}b^4)^{3/5}\right]^4=\frac{( A_{N-4}b^4)^4}
{b^{5N}}\quad \Rightarrow \quad b^{5N}=1.
\end{equation*}
Rearranging (\ref{bethe5.k}), the auxiliary functions obey the recurrences
\begin{align*}
\F_5(a)&=
\frac{(x^{24}a/b,x^{28}a/b,x^{28}a/b,x^{32}a/b;x^{2r})_{\infty}}
{(x^8a/b,x^{12}a/b,x^{12}a/b,x^{16}a/b;x^{2r})_{\infty}}
\frac{\F_5(x^{2s}a)}{\F_5(x^{4s}a)}, \nonumber \\
&&\nonumber\\
\G_5(1/a)&=
\frac{(x^{32}b/a,x^{36}b/a,x^{36}b/a,x^{40}b/a;x^{2r})_{\infty}}
{(x^{48}b/a,x^{52}b/a,x^{52}b/a,x^{56}b/a;x^{2r})_{\infty}}
\frac{\G_5(x^{2s}/a)}{\G_5(x^{4s}/a)}.
\end{align*}
The solutions are
\begin{align*}
\F_5(a)=& \frac{(x^{32}a/b,x^{40}a/b,x^{44}a/b;x^{2r})_{\infty}}
{(x^{12}a/b,x^{16}a/b,x^{24}a/b;x^{2r})_{\infty}}
\nonumber \\
&\times
\frac{(x^{28}a/b,x^{36}a/b,x^{40}a/b,x^{44}a/b,x^{48}a/b,x^{56}a/b;x^{12s})_
{\infty}}
{(x^{8}a/b,x^{12}a/b,x^{20}a/b,x^{64}a/b,x^{72}a/b,x^{76}a/b;x^{12s})_{\infty}}
,\\
\G_5(1/a)=&
\frac{(x^{40}b/a,x^{36}b/a,x^{48}b/a;x^{2r})_{\infty}}{(x^{56}b/a,x^{6 
4}b/a,x^{6
8}b/a;x^{2r})_{\infty}}
\nonumber \\
&\times
\frac{(x^{32}b/a,x^{36}b/a,x^{44}b/a,x^{88}b/a,x^{96}b/a,x^{100}b/a;x^{12s})
_{\infty}}
{(x^{52}b/a,x^{60}b/a,x^{64}b/a,x^{68}b/a,x^{72}b/a,x^{80}b/a;x^{12s}) 
_{\infty}}
,
\end{align*}
which we next substitute into the eigenvalue expression
\begin{align*}
\frac{\Lambda_5}{3}=&-\frac{w^3}{b^3}
\frac{(x^{24}w/b,x^{28}w/b,x^{36}w/b,x^{4}b/w,x^{12}b/w,x^{16}b/w;x^{2r})
_{\infty}}
{(x^{4}w/b,x^{12}w/b,x^{16}w/b,x^{24}b/w,x^{28}b/w,x^{36}b/w;x^{2r})_{\infty}}
\nonumber \\
&
\times \F_5(x^{2s}w)\G_5(1/x^{2s}w),
\end{align*}
to obtain (with $b=-1$) an expression in elliptic functions of nome $x^{12s}$
\begin{equation}
\frac{\Lambda_5}{\Lambda_0} = w^3
\frac{E(-x^{4}/w)E(-x^{12}/w)E(-x^{16}/w)E(-x^{40}w)E(-x^{48}w)E(-x^{52}w)}
{E(-x^{4}w)E(-x^{12}w)E(-x^{16}w)E(-x^{40}/w)E(-x^{48}/w)E(-x^{52}/w)}.
\label{eig5}
\end{equation}

\subsection{Mass $m_6$}
We begin the perturbation argument with
$w_j=a_j$ for $j=1, \ldots, N-5$ and $w_{N-4}=b_1x^{20}$, $w_{N-3}=b_2x^{-16}$,
$w_{N-2}=b_3x^{16}$,
$w_{N-1}=b_4x^{-8}$, $w_N=b_5x^{8}$. We can show that the $b_i$ are equal,
and we call them $b$.
The Bethe equation for the other roots is
\begin{multline}
\omega \left[a\frac{E(x^{2s}/a)}{E(x^{2s}a)}\right]^N=
(A_{N-5}b^5)^{3/5}\frac{a^5}{b^5}
\frac{E(x^{4}b/a)E(x^{8}b/a)}{E(x^4a/b)E^2(x^{8} a/b)}  \\
\times\frac{E(x^{12}b/a)E(x^{16}b/a)}
{E(x^{12}a/b)E(x^{16}a/b)}
\prod_{j=1}^{N-5} \frac{E(x^{2s}a/a_j)E(x^{4s}a_j/a)}
{E(x^{2s}a_j/a)E(x^{4s}a/a_j)}. \label{bethe6.k}
\end{multline}
In the $x \to 0$ limit this gives the equation
\begin{equation*}
a^{N-5}-\frac{1}{\omega} (A_{N-5}b^5)^{3/5}/b^5=0,
\end{equation*}
which leads in the usual way to the expression
\begin{equation*}
  \frac{1}{\omega} (A_{N-5}b^5)^{3/5}=A_{N-5}b^5.
\end{equation*}
From this and the other Bethe equations
\begin{equation*}
\left[\frac{1}{\omega} (A_{N-5}b^5)^{3/5}\right]^5=\frac{( A_{N-5}b^5)^5}
{b^{5N}}\quad \Rightarrow \quad b^{5N}=1.
\end{equation*}

After rearranging (\ref{bethe6.k}), the auxiliary functions obey the
recurrences
\begin{align*}
\F_6(a)&=
\frac{(x^{24}a/b,x^{28}a/b,x^{32}a/b,x^{36}a/b;x^{2r})_{\infty}}
{(x^4a/b,x^{8}a/b,x^{12}a/b,x^{16}a/b;x^{2r})_{\infty}}
\frac{\F_6(x^{2s}a)}{\F_6(x^{4s}a)}, \nonumber \\
\G_6(1/a)&=
\frac{(x^{28}b/a,x^{32}b/a,x^{36}b/a,x^{40}b/a;x^{2r})_{\infty}}
{(x^{48}b/a,x^{52}b/a,x^{56}b/a,x^{60}b/a;x^{2r})_{\infty}}
\frac{\G_6(x^{2s}/a)}{\G_6(x^{4s}/a)}.
\end{align*}
The solutions are
\begin{align*}
\F_6(a)=& \frac{(x^{36}a/b,x^{40}a/b;x^{2r})_{\infty}}
{(x^{16}a/b,x^{20}a/b;x^{2r})_{\infty}}\\
&\times
\frac{(x^{32}a/b,x^{36}a/b,x^{40}a/b,x^{44}a/b,x^{48}a/b,x^{52}a/b;x^{12s})
_{\infty}}
{(x^{4}a/b,x^{8}a/b,x^{12}a/b,x^{16}a/b,x^{68}a/b,x^{72}b/a;x^{12s})_{\infty}},
\\
\G_6(1/a)=&
\frac{(x^{40}b/a,x^{44}b/a;x^{2r})_{\infty}}{(x^{60}b/a,x^{64}b/a;x^{2r})_
{\infty}}\\
&\times
\frac{(x^{28}b/a,x^{32}b/a,x^{36}b/a,x^{40}b/a,x^{92}b/a,x^{96}b/a;x^{12s})_
{\infty}}
{(x^{56}b/a,x^{60}b/a,x^{64}b/a,x^{68}b/a,x^{72}b/a,x^{76}b/a;x^{12s}) 
_{\infty}}
,
\end{align*}
which we next substitute into the eigenvalue expression
\begin{equation*}
\frac{\Lambda_6}{3}=-\frac{w^3}{b^3}
             \frac{(x^{28}w/b,x^{32}w/b,x^{8}b/w,x^{12}b/w;x^{2r})_{\infty}}
{(x^{8}w/b,x^{12}w/b,x^{28}b/w,x^{32}b/w);x^{2r})_{\infty}}
\F_6(x^{2s}w)\G_6(1/x^{2s}w),
\end{equation*}
to obtain (with $b=-1$) an expression in elliptic functions of nome $x^{12s}$
\begin{equation}
\frac{\Lambda_6}{\Lambda_0} = w^3
\frac{E(-x^{8}/w)E(-x^{12}/w)E(-x^{16}/w)E(-x^{44}w)E(-x^{48}w)E(-x^{52}w)}
{E(-x^{8}w)E(-x^{12}w)E(-x^{16}w)E(-x^{44}/w)E(-x^{48}/w)E(-x^{52}/w)}.
\label{eig6}
\end{equation}

\subsection{Mass $m_7$}
We begin with
$w_j=a_j$ for $j=1, \ldots, N-6$ and  $w_{N-5}=b_1x^{-18}$,
$w_{N-4}=b_2x^{18}$,
$w_{N-3}=b_3x^{-14}$, $w_{N-2}=b_4x^{14}$, $w_{N-1}=b_5x^{-10}$,
$w_N=b_6x^{10}$. Once again the $b_i(=b)$
are all equal. The Bethe equation for the other roots is
\begin{multline}
-\omega \left[a\frac{E(x^{2s}/a)}{E(x^{2s}a)}\right]^N=
(A_{N-6}b^6)^{3/5}\frac{a^6}{b^6}
\frac{E(x^{6}b/a)E^2(x^{10}b/a)}{E(x^6a/b)E^2(x^{10} a/b)} \nonumber \\
\times\frac{E^2(x^{14}b/a)E(x^{18}b/a)}
{E^2(x^{14} a/b)E(x^{18}a/b)}
\prod_{j=1}^{N-6} \frac{E(x^{2s}a/a_j)E(x^{4s}a_j/a)}
{E(x^{2s}a_j/a)E(x^{4s}a/a_j)}. \label{bethe7.k}
\end{multline}
In the $x \to 0$ limit this gives
\begin{equation*}
a^{N-6}+\frac{1}{\omega} (A_{N-6}b^6)^{3/5}/b^6=0,
\end{equation*}
which leads to the expression in the various coefficients
\begin{equation*}
  \frac{1}{\omega} (A_{N-6}b^6)^{3/5}=A_{N-6}b^6,
\end{equation*}
and from the six Bethe equations involving $b$,
\begin{equation*}
\left[\frac{1}{\omega} (A_{N-6}b^6)^{3/5}\right]^6=\frac{( A_{N-6}b^6)^6}
{b^{4N}}\quad \Rightarrow \quad b^{4N}=1.
\end{equation*}

The recurrences to be solved for the auxiliary functions are
\begin{align*}
\F_7(a)&=
\frac{(x^{22}\frac{a}{b},x^{26}\frac{a}{b},x^{26}\frac{a}{b},x^{30}\frac{a}{b},
x^{30}\frac{a}{b},x^{34}\frac{a}{b};x^{2r})_{\infty}}
{(x^{6}\frac{a}{b},x^{10}\frac{a}{b},x^{10}\frac{a}{b},x^{14}\frac{a}{b},
x^{14}\frac{a}{b},x^{18}\frac{a}{b};x^{2r})_{\infty}}
\frac{\F_7(x^{2s}a)}{\F_7(x^{4s}a)},\\
\G_7(1/a)&=
\frac{(x^{30}\frac{b}{a},x^{34}\frac{b}{a},x^{34}\frac{b}{a},x^{38}\frac{b}{a},
x^{38}\frac{b}{a},x^{42}\frac{b}{a};x^{2r})_{\infty}}
{(x^{46}\frac{b}{a},x^{50}\frac{b}{a},x^{50}\frac{b}{a},x^{54}\frac{b} 
{a},x^{54}
\frac{b}{a},
x^{58}\frac{b}{a};x^{2r})_{\infty}}
\frac{\G_7(x^{2s}/a)}{\G_7(x^{4s}/a)},
\end{align*}
which have solution
\begin{align*}
\F_7(a)=& \frac{(x^{34}\frac{a}{b},x^{38}\frac{a}{b},x^{42}\frac{a}{b},
x^{46}\frac{a}{b};x^{2r})_{\infty}}
{(x^{10}\frac{a}{b},x^{14}\frac{a}{b},x^{18}\frac{a}{b},x^{22}\frac{a} 
{b};x^{2r}
)_{\infty}}
\nonumber \\
&\times
\frac{(x^{30}\frac{a}{b},x^{34}\frac{a}{b},x^{38}
\frac{a}{b},x^{42}\frac{a}{b},x^{42}\frac{a}{b},
x^{46}\frac{a}{b},x^{50}\frac{a}{b},x^{54}\frac{a}{b};x^{12s})_{\infty}}
{(x^{6}\frac{a}{b},x^{10}\frac{a}{b},x^{14}\frac{a}{b},
x^{18}\frac{a}{b},x^{66}\frac{a}{b},x^{70}\frac{a}{b},x^{74}\frac{a}{b},
x^{78}\frac{a}{b};x^{12s})_{\infty}},
\end{align*}
\begin{align*}
\G_7(1/a)=&
\frac{(x^{34}\frac{b}{a},x^{38}\frac{b}{a},x^{42}\frac{b}{a},x^{46}\frac{b}{a};x
^{2r})_{\infty}}
{(x^{58}\frac{b}{a},x^{62}\frac{b}{a},x^{66}\frac{b}{a},x^{70}\frac{b} 
{a};x^{2r}
)_{\infty}}
\nonumber \\
&\times
\frac{(x^{30}\frac{b}{a},x^{34}\frac{b}{a},x^{38}
\frac{b}{a},x^{42}\frac{b}{a},x^{90}\frac{b}{a},x^{94}
\frac{b}{a},x^{98}\frac{b}{a},x^{102}\frac{b}{a};x^{12s})_{\infty}}
{(x^{54}\frac{b}{a},x^{58}\frac{b}{a},x^{62}\frac{b}{a},
x^{66}\frac{b}{a},x^{66}\frac{b}{a},x^{70}
\frac{b}{a},x^{74}\frac{b}{a},x^{78}\frac{b}{a};x^{12s})_{\infty}}.
\end{align*}
Substitution into
\begin{align*}
\frac{\Lambda_7}{3}=&\frac{w^4}{b^4}\frac{(x^{22}\frac{w}{b},x^{26}
\frac{w}{b},x
^{30}\frac{w}{b},x^{34}\frac{w}{b},
x^{6}\frac{b}{w},x^{10}\frac{b}{w},x^{14}\frac{b}{w},x^{18}\frac{b}{w};x^{2r})
_{\infty}}
{(x^{6}\frac{w}{b},x^{10}\frac{w}{b},x^{14}\frac{w}{b},x^{18}\frac{w}{b},
x^{22}\frac{b}{w},x^{26}\frac{b}{w},x^{30}\frac{b}{w},x^{34}\frac{b}{w };x^{2r})_
\infty}
\nonumber
\\
& \times\F_7(x^{2s}w)\G_7(1/x^{2s}w),
\end{align*}
yields the result (with $b=-1$ and elliptic nome $x^{12s}$)
\begin{multline}
\frac{\Lambda_7}{\Lambda_0} = w^4
\frac{E(-x^{6}/w)E(-x^{10}/w)E(-x^{14}/w) E(-x^{18}/w)}
      {E(-x^{6}w)E(-x^{10}w)E(-x^{14}w) E(-x^{18}w)}
\\
\times \frac{E(-x^{42}w)E(-x^{46}w)
E(-x^{50}w)E(-x^{54}w)}
{E(-x^{42}/w)E(-x^{46}/w)E(-x^{50}/w)
E(-x^{54}/w)} . \label{eig7}
\end{multline}

\newpage

\begin{table}[h]
\caption{The integers appearing in (\ref{m8}) and (\ref{prop}) for $L=3$.}
\label{eight}
\begin{center}
\begin{tabular}{cl}
\hline \hline
$j$ &   $a$ \\
\hline
1   &        1, 11 \\
2   &        7, 13 \\
3   &        2, 10, 12 \\
4   &        6, 10, 14 \\
5   &        3, 9, 11, 13 \\
6   &        6, 8, 12, 14\\
7   &        4, 8, 10, 12, 14\\
8   &        5, 7, 9, 11, 13, 15 \\
\hline \hline
\end{tabular}
\end{center}
\end{table}


\begin{table}[ht]
\caption{The integers appearing in (\ref{prop}) for $L=4$.}
\label{seven}
\begin{center}
\begin{tabular}{cl}
\hline \hline
$j$ &   $a$ \\
\hline
1&       6 \\
2&       1, 7 \\
3&       4, 8 \\
4&       5, 7 \\
5&       2, 6, 8 \\
6&       4, 6, 8 \\
7&       3, 5, 7, 9 \\
\hline \hline
\end{tabular}

\end{center}
\end{table}


\begin{table}[ht]
\caption{The integers appearing in (\ref{prop}) for $L=6$.}
\label{six}
\begin{center}
\begin{tabular}{cl}
\hline \hline
$j$ &  $a$ \\
\hline
1 ,$\bar 1$&       4 \\
2&      1, 5 \\
3, $\bar 3$&    3, 5 \\
4&     2, 4, 6 \\
\hline \hline
\end{tabular}
\end{center}
\end{table}


\begin{table}[ht]
\caption{String positions $u_j$ and corresponding eigenvalue bands
for the seven elementary mass excitations $m_i$ of the dilute $A_4$ model
in regime 2$^-$ \cite{GNp}. The strings are in units of $\pi /20$.}
\label{strings}
\begin{center}
\begin{tabular}{cll}
\hline \hline
$i$ &  String positions& Band \\
\hline
1   &       $\pm 2, 10$& $w^{}$  \\
2   &       $\pm 7$ &$w^2$\\
3   &       $\pm 6, 10$&$w^2$ \\
4   &       $\pm 3, \pm 9$ &$w^2$  \\
5   &       $\pm 6, \pm 8$&$w^3$  \\
6   &       $\pm 4, \pm 8, 10$ &$w^3$\\
7   &       $\pm 5, \pm 7, \pm 9$ &$w^4$\\
\hline \hline
\end{tabular}
\end{center}
\end{table}

\begin{table}[ht]
\caption{The four perturbations of the tricritical Ising model, and the
objects from statistical mechanics to which they are related in the
scaling limit.}
\label{TIM}
\begin{center}
\begin{tabular}{lcclc}
\hline \hline
Perturbation&Field&Weight&IRF  model&$\mathcal{H}_{\rm BEG}$\\
\hline
&&&\\
Leading magnetic & $\phi_{(2,2)}$ & $\frac{3}{80}$ &  Not
integrable& $H$\\
&&&\\
Leading thermal & $\phi_{(1,2)}$ & $\frac{1}{10}$ &  Dilute
A$_4$, regime 2& $1/J$\\
&&&\\
Subleading magnetic & $\phi_{(2,1)}$ & $\frac{7}{16}$ &  Dilute
A$_3$, regime 1& $H_3$\\
&&&\\
Subleading thermal & $\phi_{(1,3)}$ & $\frac{3}{5}$ &  ABF A$_4$,
regime III&$D$\\&&&\\
\hline \hline
\end{tabular}
\end{center}
\end{table}

\end{document}